\documentclass[acmsmall,nonacm]{acmart}

\usepackage{graphicx}
\usepackage{tikz}
\usepackage[utf8]{inputenc}
\usepackage{pgfplots}
\usepackage{pifont}
\usepackage{microtype}
\usepackage{cprotect}
\usepackage[section]{placeins}

\usepackage{listings}
\lstdefinelanguage{sparql}
{
  morekeywords={SELECT,OPTIONAL,FROM,DISTINCT,WHERE,FILTER,GROUP,ORDER,LIMIT,BY,IN,AS,COUNT},
  sensitive=true, 
  morecomment=[l]{//}, 
  morecomment=[s]{/*}{*/}, 
  morestring=[b]" 
}

\AtBeginDocument{%
  \providecommand\BibTeX{{%
    \normalfont B\kern-0.5em{\scshape i\kern-0.25em b}\kern-0.8em\TeX}}}

\acmJournal{JDIQ}

\begin{document}
\title{Open Data and the Status Quo - A Fine-Grained Evaluation Framework for Open Data Quality and an Analysis of Open Data portals in Germany}

\author{Lisa Wenige}
\author{Claus Stadler}
\author{Michael Martin}
\email{{surname}@infai.org}
\affiliation{%
  \institution{Institute for Applied Informatics}
  \streetaddress{Goerdelerring 9}
  \city{Leipzig}
  \state{Saxony}
  \country{Germany}
  \postcode{04109}
}

\author{Richard Figura}
\author{Robert Sauter}
\author{Christopher W. Frank}
\email{{r|r|c.surname}@ciss.de}
\affiliation{%
  \institution{CISS TDI GmbH}
  \streetaddress{Barbarossastraße 36}
  \city{Sinzig}
  \state{Rhineland Palatinate}
  \country{Germany}
  \postcode{53489}
}

\renewcommand{\shortauthors}{Wenige and Stadler, et al.}

\begin{abstract}
This paper presents a framework for assessing data and metadata quality within Open Data portals. Although a few benchmark frameworks already exist for this purpose, they are not yet detailed enough in both breadth and depth to make valid statements about the actual discoverability and accessibility of publicly available data collections. To address this research gap, we have designed a quality framework that is able to evaluate data quality in Open Data portals on dedicated and fine-grained dimensions, such as interoperability, findability, uniqueness or completeness. Additionally, we propose quality measures that allow for valid assessments regarding cross-portal findability and uniqueness of dataset descriptions. We have validated our novel quality framework for the German Open Data landscape and found out that metadata often still lacks meaningful descriptions and is not yet extensively connected to the Semantic Web.
\end{abstract}



\maketitle

\section{Introduction}
Open Data has great potential for society as users and organizations can gain helpful insights from analyzing publicly available datasets and might also create valuable data-driven applications based on these data sources \cite{benefits}. Open Data portals (ODP) can also help policy-makers to obtain access to data in order to tackle complex societal issues \cite{arzberger}. For more than two decades, the Open Data movement has gained momentum around the world. Open Data portals have been set up by governmental bodies at the provincial, regional or national level \cite{recommendations}. According to the latest edition of the Open Data Barometer \cite{odb}, more than half of the 115 surveyed countries in the study had an official Open Data initiative in place to foster open publication of datasets. These infrastructures may form the backbone of our information society as they can connect data owners with users.\\
However, only a well-equipped Open Data landscape can really foster availability and findability of data sources. Incomplete or inaccessible metadata descriptions, on the other hand, can hinder wide adoption and application of valuable data. For this purpose, Open Data portals need to provide adequate organizational and technical surroundings as well as high quality metadata descriptions. Although there is agreement that the quality of the data descriptions is crucial for the findability and thus also for the widespread use of the data, there are so far only a few systematic approaches as to how Open Data quality can be measured comprehensively and scientifically.\\
Among the existing approaches, the focus often either solely lies on the completeness of metadata descriptions or on the governmental policies that are in place to foster an Open Data environment \cite{knippenberg}. What is missing, however, is the definition of a comprehensive set of qualitative as well as quantitative criteria that can be applied to measure the overall quality of the metadata descriptions in Open Data portals, e.g. in terms of accessibility, uniqueness, legal security or completeness. In addition to that, the aspect of data quality is mostly investigated by surveying policymakers and practitioners, rather than evaluated on the data itself. Thus, the meaningfulness of the results might be diluted by self-reported data \cite{recommendations}.\\
We present a framework for Open Data quality evaluation that is intended for direct application on data and metadata descriptions in Open Data portals. Our framework is able to systematically capture the current situation of Open Data quality in multiple dimensions. This enables benchmarking and comparison of different regions, countries and thematic areas with a stable set of quality metrics. Hence, politicians and practitioners can derive appropriate courses of action by utilizing our framework. For such an evaluation framework to be widely applicable, it also needs to define the respective units of analysis. Even though it is often interesting to analyze the quality of a single Open Data portal, most of the benchmarks have a broader perspective, especially in cases where the analysis informs policy makers and practitioners in order to optimize existing infrastructures. For this purpose, we propose the notion of Open Data landscapes. It describes sets of Open Data portals that can be grouped together because of their {\itshape geographic, administrative or thematic} proximity and that should be therefore analyzed simultaneously. This paper presents a comprehensive plan of action for Open Data quality evaluation and is structured as follows: 
\begin{itemize}
\item \textbf{Section 2} gives an overview of the state of the art in Open Data quality evaluation. 
\item \textbf{Section 3} defines the term \textit{Open Data landscape} as a potential unit of analysis for Open Data quality evaluation and categorizes potential types of \textit{Open Data landscapes}.
\item \textbf{Section 4} presents the individual dimensions of our Open Data quality evaluation framework and defines concrete quality metrics and procedures with which quality can be measured.
\item \textbf{Section 5} applies the evaluation framework to the use case of the German \textit{Open Data landscape} thus validating the framework's applicability. It presents the respective results with regard to Open Data quality in this country
\item \textbf{Section 6} discusses the most important findings and proposes future research directions. 
\end{itemize}

\section{Related Work}
Given the high prevalence of Open Data as well as the growing importance of dataset search engines (e.g., \textit{Google Dataset Search} \cite{brickley}) that heavily depend on comprehensive dataset descriptions, it is astonishing that only little systematic research has been conducted in order to better assess the data quality in Open Data landscapes \cite{benefits}. Since data quality is an important precondition for data re-use, it is surprising that the vast majority of Open Data studies and benchmarks has not put a strong focus on this aspect or at least does not propose concrete metrics and subsequent steps for its improvement. For instance, Suscha et al. conducted a meta-analysis of the five most important Open Data benchmarks and note that \cite{susha}:

\begin{quote} \textit{``Most open data benchmarks produce results that are generic or ambiguous for any particular organization.''}\end{quote}

This finding underpins the need for comprehensive frameworks to measure Open Data quality.
While a good amount of studies on data statistics as well as quality issues has already been done in the field of Linked Open Data and Semantic Web \cite{lodstats,zaveri,schmachtenberg}, the same can not be said about Open Data publishing in general. Many of the published papers showcase examples of Open Data application \cite{napoli,maritime,chem2bio}, present concepts and theories for public data sources in general \cite{bertot,mcdermott,zeleti} or propose certain architectures or technologies for easier publication and consumption of datasets \cite{datagraft,architectural,stadler}.\\
Some generic Open Data benchmarks only touch the aspect of data quality marginally. The \textit{2020 Open Data Barometer} determines the advancement of Open Data in different European countries based on the four pillars: \textit{Policy}, \textit{Portals}, \textit{Impact} and \textit{Quality}. Even though data quality is investigated in this research, it is only assessed based on the aspects of data completeness and freshness. Additionally, this study relies on self-reported data (i.e., answers by Open Data practitioners and policy makers) and does therefore not include evaluation protocols and results for the data sources themselves \cite{odb}. Another paper by Veljković et al. proposes to benchmark Open Data portals alongside the dimensions \textit{Data Openness, Transparency, Participation,} and \textit{Collaboration} \cite{benchmarking}. While this is a helpful approach for assessing the maturity of Open Data portals it falls short when data quality has to be evaluated comprehensively. Similar performance indicators are mentioned in the Open Data benchmark framework proposed by Sayogo et al., even though the authors do not provide the exact measures to capture performance in the individual dimensions \cite{sayogo}.\\
Little research has been conducted on the systematic measurement of Open Data quality. Sasse et al. note that there is still a lack of viable qualitative and quantitative indicators that can be applied to assess the performance of data quality and reuse \cite{recommendations}. Among the few existing approaches, the one by Braunschweig et al. strikes as quite advanced. The authors have surveyed Open Data portals and analyzed the number of datasets, file formats and whether the platforms provide an API. However, only five Open Data portals have been comprehensively analyzed with regard to the quality of their metadata descriptions \cite{braunschweig}. While the approach taken by Braunschweig is promising, it should be based on an extensive evaluation framework that defines quality metrics as well as is carried out for a larger number of Open Data portals in order to gain extensive insights. Neumaier et al., on the other hand, performed such an extensive analysis. The authors crawled hundreds of Open Data portals and analyzed the metadata descriptions that were attached to the datasets. However, despite their comprehensive data collection, the approach mostly captures the existence of certain metadata fields \cite{neumaier}. In contrast, the study by Kubler et al. goes further in that it defines a set of performance indicators, such as openness or DCAT conformance \cite{ahp}. Among the surveyed papers, this evaluation framework is the most comprehensive one. However, even in this approach the measurements could be even more detailed for certain quality aspects. In this line of research, Kiraly and Büchler propose some additional quality metrics for bibliographic data, such as the \textit{uniqueness} score \cite{kiraly} that could be well applied for metadata descriptions of ODPs.\\
This work strives to advance the existing approaches in order to provide a comprehensive framework for Open Data quality evaluation. 

\section{Open Data Portals \& Open Data Landscapes}
With our framework, We aim to assess the extent to which suitable technologies and metadata descriptions are already being used to make Open Data collections available, findable and usable for a broad public. The framework is applicable to Open Data portals and Open Data landscapes. The term {\itshape Open Data Landscapes} (ODL) describes a {\itshape geographic, administrative or thematic space} that is comprised of several Open Data portals. The ODL categories have the following characteristics:
\begin{itemize}
\item {\itshape Geographic:} A geographic landscape can be a larger (e.g., Africa) or a smaller (e.g. Scandinavia) territorial area that the portals in question cover through their data collections. 
\item {\itshape Administrative:} This type of ODL is characterized by data portals that have been set up by public bodies at the same municipal, state or federal level and predominantly contain data that has been gathered within the scope of responsibility of the respective public body (e.g. Germany). 
\item {\itshape Thematic:} This category describes portal landscapes that address a specific topic (such as biodiversity or neural networks). Determining the status quo of such thematically focused landscapes is highly relevant, especially in research data management, since such analyses can help to decide whether a particular thematic area is already sufficiently equipped with Open Data technologies.
\end{itemize}
We assume that using ODLs as a possible unit of quality evaluation is suitable to grasp the sophistication of Open Data portals in a certain area. In contrast, an exclusive analysis of individual portals may sometimes not be sufficient since many phenomena (e.g. replication of datasets or uniqueness of identifiers across platforms) can only be grasped by a cross-portal analysis. Moreover, in many cases, users are more interested in the retrievability of the actual data collections, rather than their provenance. Thus, an analysis at ODL level can serve as the starting point for subsequent measures that create an impact for the entire Open Data landscape under consideration and therefore helps to better connect the data with its users. On the other hand, an analysis of a single ODP can be interesting as well. From the point of view of a platform provider it is still helpful to be able to identify optimization potential locally. Besides its applicability to ODPs and ODLs alike, our framework can also be used to assess the status quo of Open Data portals worldwide. However, such an analysis would then be more exploratory in nature, as it would not fall within the scope of responsibility of any organization to apply concrete policy measures to improve the status quo.\\

\section{Evaluation Model}
\begin{figure}[h!]
    \centering
   \includegraphics[width=0.95\textwidth]{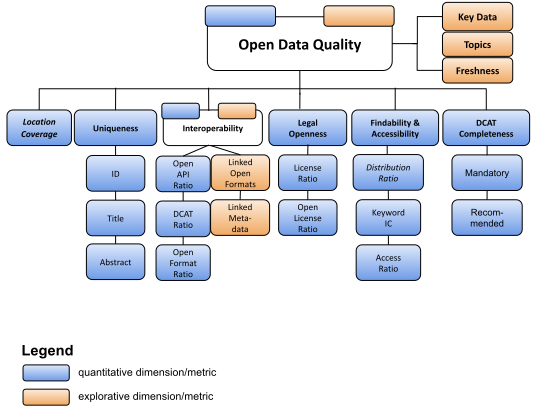}
    \caption{Overview of the Open Data evaluation framework}
    \label{fig:my_label}
\end{figure}

This section presents the Open Data evaluation framework with its multiple quality criteria and dimensions.
The metadata descriptions of Open Data collections in DCAT format serve as the data basis of the framework. DCAT is a W3C recommendation and defines an RDF-based format that enables data owners to describe data collections in a uniform way and thus ensures the interoperability of data catalogs on the web. DCAT can be considered the de-facto standard for Open Data publishing \cite{dcat}. Numerous data portals offer a DCAT interface with which data descriptions can be read automatically. Where this is not possible, a transformation into DCAT format based on an appropriate mapping (see Sect. \ref{sec:usecase}) is recommended so that the analysis can be performed on a common data basis.\\
Figure \ref{fig:my_label} depicts the explorative (orange) and quantitative (blue) dimensions (squares with bold letters) and subdimensions (squares with regular letters) of our framework. The aim of the framework is to comprehensively evaluate the quality of the data and metadata contained in Open Data portals. This is because only sufficient meta-information and a technologically open infrastructure will enable stakeholders to actually find and use the data sources contained in portals. We developed this framework such that it can be applied to assess individual Open Data portals. With the help of our approach, it is also possible to evaluate Open Data landscapes as well as to compare ODPs or ODLs within a benchmark where this is necessary. Evaluations should help policymakers and portal providers alike to derive actionable recommendations to optimize their technical solutions. 
In the event that ODLs are evaluated, we incorporated two cross-portal performance indicators (marked in italics: \textit{Location Coverage} and \textit{Distribution Ratio}) into our set of metrics in order to increase the understanding of the landscape as a whole However, in the event that only one portal is currently being checked, these metrics can be omitted without loss of information for the remaining indicators.\\
On of the key strengths of our approach is that its set of metrics is more comprehensive than that of other evaluation frameworks. These concrete quality metrics can be applied to quantify abstract concepts, such as \textit{Uniqueness}, \textit{Findability} and \textit{DCAT Completeness}. The metrics are formally defined in detail in the following subsections. However, although the framework is quantitatively oriented, it also comprises qualitative dimensions. This helps to gain a holistic understanding of the Open Data portals that are being investigated. For benchmark purposes, individual ODPs or ODLs can be ranked according to the \textit{Analytic Hierarchy Process} (AHP) as proposed by Kubler et al. With these techniques, the portals or landscapes can be compared in terms of their performance (i.e., score values) within the quantitative subdimensions. These scores are then aggregated using individual weights for each dimension to produce the final ranking score of the ODP or ODL \cite{ahp}. However, in the event that only a single ODP or ODL is to be evaluated, the framework can be used standalone instead. The following subsections describe the need and the procedures for measuring each performance indicator in detail. 

\subsection{Key Data}
In this category, it is determined in which orders of magnitude data collections are made available. Questions, such as how many datasets or files are registered can help to assess whether the quantity of datasets appropriately reflects the size and importance of the Open Data portal(s) under consideration. The key data can be obtained by issuing the following SPARQL query to the metadata catalog that comprises DCAT descriptions.
\newpage
\begin{lstlisting}[xleftmargin=.225\textwidth, xrightmargin=.225\textwidth]
prefix dcat: <http://www.w3.org/ns/dcat#> 

SELECT COUNT(DISTINCT ?ds) as ?ds COUNT(DISTINCT ?dist) as ?dist   COUNT(DISTINCT ?url) as ?url WHERE {

?catalog dcat:dataset ?ds .
?ds dcat:distribution ?dist  .
?dist dcat:accessURL ?url .
FILTER (!isBlank(?ds))
FILTER (!isBlank(?dist))

}
\end{lstlisting}

\subsection{Topics}
To this date, systematic content analyses are uncommon in Open Data benchmarks. A rare example is the survey done by Thorsby et al.  who identified frequently occurring dataset topics being listed in ODPs of American cities. This analysis was guided by the list of subject categories provided by the respective Open Data portal \cite{thorsby}. However, such a list may be incomplete in that it does not contain the keywords that match the data collections. Second, not every ODP has such a category list. Therefore, conducting an exploratory topic analysis using natural language processing (nlp) techniques is advisable in any case.\\ The identification of the most frequent topics can reveal whether a data landscape is sufficiently diverse or focuses rather one-sidedly on one or two main topics, whereby other equally important topics tend to be neglected. To perform a meaningful topic analysis, the corresponding text passages of the metadata should be evaluated. For this purpose, it is useful to analyze the title, the description or the assigned keywords of the datasets in question. In addition, established state-of-the-art procedures for text analysis and topic modeling, such as \textit{Latent Semantic Analysis} or \text{Latent Dirichlet Allocation} can be applied for this purpose. These are unsupervised machine learning techniques that can explore and discover topics in documents based on the analysis of frequently occurring words in documents \cite{dumais, blei}. 

\subsection{Freshness}
\label{subsec:fresh}
Open Data portals should provide fresh data sources since end users depend on current information in order derive meaningful conclusions or implement helpful data-driven services based on these data collections. However, an overall evaluation of the average timeliness of datasets being registered in an ODP or ODL is hardly possible without the knowledge of the needs of end users. Thus, the assessment of the timeliness of a data collection is always relative to a particular task. Corresponding metrics measure the difference of the user's desired freshness (e.g. up date intervals of one day) in relation to the actual update intervals \cite{zaveri}. However, such an approach would require detailed and specific knowledge of individual user goals which might in turn differ tremendously between different users.\\
In this way, a fine-grained survey of user needs is difficult and should ideally be determined in a separate survey. For a generic assessment of Open Data landscapes - which is the scope of our evaluation framework - the performance dimension of freshness can only be captured through an explorative analysis. Once the specific user needs regarding timeliness are known for the datasets in a particular ODP, freshness can be directly measured and this dimension can be turned into a quantitative one. 
For a pure explorative analysis, the distribution of update intervals is measured by the difference between the date of the data crawl and the timestamp of the last modification of the dataset (i.e., as determined by the metadata feature \verb+dcat:modified+).

\subsection{Location Coverage}
This quality indicator only applies to benchmarks being conducted on Open Data landscapes with a {\itshape geographic} or {\itshape administrative} focus. Data portals located in such an ODL can usually be assigned to units of a specific location code scheme (LOCO). Typical location code schemes are ISO 13166, UN/LOCODE or NUTS. The latter one has been defined for regions within the European Union. LOCO schemes typically describe a hierarchical structure for administrative units. For instance, the NUTS scheme defines four hierarchy levels. {\itshape NUTS 0} denotes nation states, {\itshape NUTS 1} major socio-economic regions, while {\itshape NUTS 2} and {\itshape NUTS 3} describe medium and small regions respectively. Many ODPs can be mapped to a location level. The location coverage for a certain level of a LOCO scheme measures the proportion of administrative units that are covered by ODPs. A high location coverage is a positive quality indicator since it shows that datasets of the majority of administrative units are registered in respective data portals.
\begin{equation}LOCO{\text -}Ratio_{level} = \frac{\# \; ODPs \; at \; level}{ \# \; regions \; at \; level}\end{equation} 

\subsection{Uniqueness}
Another quality indicator when considering Open Data quality is the aspect of metadata uniqueness. Each dataset should have a unique identifier, title or description. This helps potential users to detect useful datasets that meet their information needs and helps to distinguish data collections from others. Unique descriptions are also vital for helping search engine crawlers to find and unambiguously index datasets. For instance, the developers guide of the \textit{Google Dataset Search} strongly recommends using meaningful titles for a data collection. Instead of naming two different datasets with the same title, the guide proposes to apply more descriptive titles which can be distinguished from each other.\footnote{\url{https://developers.google.com/search/docs/data-types/dataset}}\\
Measuring uniqueness, we apply a slightly modified version of a metric proposed by Kiraly et al. which is in itself an adaptation of Solr's relevance score \cite{kiraly}. The score of a property value (e.g. the \verb+dct:title+ {\itshape Report}) which characterizes a dataset $ds$ is determined by counting the number of occurrences of this value ($v_f$) in the catalog as well as the number of datasets ($t_f$) being linked to the corresponding property $t$ (e.g., \verb+dct:title+). The set of considered properties is defined as $T$ = $\{$\verb+dct:identifier+, \verb+dct:title+, \verb+dct:description+ $\}$.\\
The counted values are applied to the score functions shown below. The properties that are connected to a dataset are denoted as $Prop_{ds}$. The final $uniqueness_f$ score is determined by relating the current score, which depends on the frequency of a property value to the ideal score that has a frequency of 1. The higher the final score, the better.
\begin{equation}
\label{eq:score}
score(ds,t_f,v_f) = 
            log\left(1+\frac{t_f-v_f+0.5}{v_f+0.5}\right), \quad t_f \in T \wedge t_f \in Prop_{ds}\\
\end{equation}
\begin{equation}
\label{eq:unique} 
uniqueness_t = \frac{score(ds,t_f,v_f)}{score(ds,t_f,1.0)}
\end{equation}

\subsection{Interoperability}
\label{subsec:inter}
In this quality dimension, it is investigated how interoperability is achieved within the ODP or ODL. Interoperability is an important factor since it facilitates cross-portal data exchange and reuse of datasets for different application domains. The following subdimensions are to be explored.
\begin{itemize}
\item {\itshape Open ratio (higher is better)}: This measures the degree to which the ODL is already covered with ODPs that provide access to their portal via an open API: 
\begin{equation}Open{\text -}Ratio = \frac{\# \; open \; API \; ODPs}{ \# \; all \; ODPs}\end{equation} 
An ODP is attributed an open API when it is possible to programmatically collect metadata for data collections in JSON or DCAT formats from the portal. In case, a portal solely provides JSON metadata, these descriptions should at least adhere to the CKAN/ DKAN metadata model, since this structure has already been successfully mapped to DCAT fields. Both CKAN/ DKAN and DCAT metadata can be regarded as de-facto standards for dataset descriptions which can be processed by a couple of publicly available software clients. 
ODPs that provide metadata in other formats, such as raw XML or RSS can not be mapped without additional efforts and often occur in custom formats which complicates the processing with crawling software. Thus, portals only providing non-standard metadata are not regarded as open API ODPs. 
\item {\itshape DCAT ratio (higher is better)} The subset of ODPs offering their dataset descriptions in DCAT formats is measured with the following ratio.
\begin{equation}DCAT{\text -}Ratio = \frac{\# \; ODPs \; with \; DCAT \; profiles}{ \# \; all \; ODPs}\end{equation} 
Generally, ODPs that provide genuine DCAT metadata can be regarded as fulfilling the highest standards in terms of metadata. The DCAT model is a W3C recommendation for describing datasets and data catalogs. Because of its design as an RDF vocabulary, it enables data integration and exchange
among public data catalogs on the web \cite{dcat}. If an ODP offers a DCAT API, it will be easier to download and save its metadata descriptions to a triple store that can be queried with SPARQL right away.
\item {\itshape Open format ratio (higher is better)}: According to the Open Definition, publicly available datasets should be provided in an open format. Open formats are those whose use is not hindered by any technical or monetary restrictions and which can be processed with at least one open-source tool \cite{open-def}. The Open format ratio is measured as the number of open formats attached to DCAT distributions among all available formats in the ODL. 
\begin{equation}Open{\text -}Format{\text -}Ratio = \frac{\# \; open \; format \; datasets }{ \# \; all \; datasets}\end{equation} 
\item {\itshape Linked Open formats (explorative)} While data collections should be described with DCAT, the actual data collections are available in different formats. Depending on the thematic focus, quite different file types fit the data. Hence, a high prevalence of typical Linked Data formats, such as \verb+rdf/xml+, \verb+turtle+ or \verb+json-ld+ is not a positive quality indicator in itself since some data types (e.g., sensor or visual data) can be more easily and efficiently represented in other formats. At the very least, however, it is a positive sign if part of the data is available in RDF as this is associated with increased interoperability. 
\item {\itshape Linked Metadata (explorative)}: This dimension explores the extent to which the provided metadata features of ODP datasets exploit the possibilities of data integration via Linked Data. Identity links (such as \verb+owl:sameAs+, \verb+skos:exactMatch+) can be used to connect items with the web of data. This is helpful when data collections are tagged with keywords that can also be found in general purpose knowledge bases, such as DBpedia or Wikidata. In this way, users receive valuable additional information about the thematic focus of the data. In addition, it can be helpful to use frequently used vocabularies of the Web of Data to describe data collections. For example, geographic locations or thematic references can be more easily identified. To get an overview of the extent of Linked Data integration within the metadata descriptions it is useful to depict the distribution of occurring namespaces. This allows us to estimate whether there are other namespaces in addition to those commonly used in DCAT catalogs (i.e., \verb+dcat+, \verb+dct+ or \verb+foaf+). A SPARQL query that extracts the required data is available on Github.\footnote{\url{https://github.com/mclient-project/crawling/blob/main/interoperability/namespaces.sparql}}
\end{itemize}
\subsection{Legal Security \& Openness} 
According to the DCAT standard, license information should be attached to the distributions of a dataset \cite{dcat}. Hence, the following metrics for legal openness are to be applied to metadata descriptions of datasets and their distributions. 
\begin{itemize}
\item {\itshape License Ratio (higher is better):} 
The license ratio measures the amount of distributions that are linked to a license document. This quality metric is vital in order to assess the legal security of platforms that provide Open Data. For each dataset a license should be clearly identifiable with the license conditions being at least publicly available in text format and at best in structured form. 
\begin{equation}License{\text -}Ratio = \frac{\# \; licensed \; distributions }{ \# \; all \; distributions}\end{equation} 
\item {\itshape Open License Ratio (higher is better):} 
The Open License ratio measures the amount of data collections that are provided with an open license. According to the \textit{Open Data Handbook} open licenses allow datasets to be ``{\itshape freely used, re-used and redistributed by anyone – subject only, at most, to the requirement to attribute and share-alike}'' \cite{handbook}. In order to identify these open licenses, we follow the list of licenses that was curated by the central German Open Data hub \verb+govdata+ \footnote{\url{https://www.dcat-ap.de/def/licenses/}}.
\begin{equation}Open{\text -}License{\text -}Ratio = \frac{\# \; distributions \; with \; an \; open \; license }{ \# \; all \; distributions}\end{equation} 
\end{itemize}

\subsection{Findability \& Accessibility}
A vital condition for wide adoption and usage of Open Data is its discoverability and usability. Therefore, options for easy retrieval should be available and dataset downloads should not be prevented by technical barriers. The following three performance indicators constitute the quality dimension of findability and accessibility.
\label{subsec:find}
\begin{itemize}
\item {\itshape Replica-Ratio (higher is better)}: This performance indicator measures the ratio of datasets that occur in different ODPs among the total number of datasets in an Open Data landscape. 
\begin{equation}Replica{\text -}Ratio = \frac{\# \; datasets \; in \; different \; ODPs }{ \# \; all \; datasets}\end{equation} 
The ratio of dataset duplicates is determined by identifying DCAT access URLs that are linked to datasets and that appear in different catalogs within an Open Data landscape. In general, it can be assumed that the retrievability increases with an increasing number of datasets that can be reached via different Open Data platforms.
\newpage
\begin{lstlisting}[xleftmargin=.25\textwidth, xrightmargin=.25\textwidth]
PREFIX dcat: <http://www.w3.org/ns/dcat#> 

SELECT COUNT(DISTINCT ?ds) WHERE {
?catalog dcat:dataset ?ds .
?ds dcat:distribution ?dist  .
?dist dcat:accessURL ?url .
FILTER (!isBlank(?ds))
FILTER (!isBlank(?dist)){
    SELECT DISTINCT ?url WHERE {
    ?catalog dcat:dataset ?dataset .
    ?catalog2 dcat:dataset ?dataset .
    ?dataset dcat:distribution ?distribution .
    ?distribution dcat:accessURL ?url 
    FILTER (?catalog != ?catalog2) 
    FILTER (!isBlank(?dataset))
    FILTER (!isBlank(?distribution)) }}
}
\end{lstlisting}
\item {\itshape Keyword Information Content (higher is better)}: Another important factor for retrieving suitable datasets are meaningful keywords that can help to identify relevant content. Keywords are usually among the item features stored first in a search index to provide appropriate retrieval access to data collections. For this reason, the keywords should be meaningful enough to make it easier to find the data  and assess its content. We propose to apply Shannon's information content function for this purpose. For each dataset $ds$, we calculate the combined information content of the set of keywords $K_{ds}$ attached to it via the property \verb+dcat:keyword+. Let $freq(k)$ be the frequency of the keyword $k$ in the entire ODL catalog $C$, $n$ be the maximum frequency of a keyword in this catalog and $card(K_{ds})$ be the number of keywords that are linked to a dataset, then the {\itshape Keyword Information Content} $IC(K_{ds})$ is defined as follows:
\begin{equation}IC(K_{ds})=
\label{eq:score}
score(t_f,v_f) = 
\left\{ 
        \begin{array}{ll} 
            - \sum_{k \in K_{ds}} \frac{1}{card(K_{ds})} \times log \frac{freq(k)}{n}, \quad K_{ds} \neq \emptyset\\
            0 , \quad \quad \quad \quad \quad \quad \quad \quad \quad \quad \quad \quad \quad  \quad  otherwise
        \end{array}
 \right.
\end{equation} 
For the sake of readability and comparability, $IC(K_{ds})$ scores should fall within the range of 0 to 1. Since the values of the information content function have a different range, a corresponding normalization must be carried out. The normalization step smooths the IC score of a particular data collection ($ds$) using the maximum and minimum scores that occur among all data collections in the catalog ($c \in C$). 
\begin{equation}
IC_{norm}(K_{ds})=\frac{IC(K_{ds})-min_{c \in C}IC(K_c)}{max_{c \in C}IC(K_c)-min_{c \in C}IC(K_c)}
\end{equation}
\item {\itshape Accessibility (higher is better)}: This metric is used to determine the extent to which the data collections can be downloaded using the $dcat:accessURL$s specified in the metadata. \verb+Curl+ requests are applied to check the accessibility and availability of the raw data files. The metric provides an overview over the extent of data that is actually publicly available. The ratio determines the amount of DCAT distributions with working accessURLs in relation to the total number of distributions in the catalog. 
\begin{equation}Accessibility{\text -}Ratio = \frac{\# \; distributions \; with \; working \; accessURLs}{\# \; distributions}\end{equation}
\end{itemize}

\subsection{DCAT Completeness}
As a quality indicator, metadata completeness is relevant because it captures the extent to which data collections are adequately described. The more comprehensively datasets are characterized, the better they can be found by users and software agents. Complete descriptions also improve the extent to which the content of the actual data collections can be assessed by a potential user.\\
The W3C Recommendation for the DCAT vocabulary defines possible fields to describe data collections. The corresponding \textit{DCAT Application Profile} by the European Union specifies this schema \cite{dcat-ap}. It defines mandatory and recommended fields thereby creating the basis for checking the completeness of data descriptions. Table \ref{tab:dcat} lists the respective properties for each category. 
\begin{table*}
\centering
\label{tab:survey}
\begin{tabular}{cc}
\toprule
\textbf{Mandatory (Mand.) DCAT fields} & \textbf{Recommended (Rec.) DCAT fields} \\
\midrule
\verb+dct:description+ & \verb+dcat:distribution+ \\
\verb+dct:title+ & \verb+dcat:keyword+ \\
& \verb+dct:temporal+ \\
& \verb+dcat:contactPoint+ \\
& \verb+dct:spatial+ \\
& \verb+dct:publisher+ \\
\bottomrule
\end{tabular}
\caption{List of mandatory and recommended DCAT fields}
\label{tab:dcat}
\end{table*}
The properties can be queried with a corresponding SPARQL query in order to obtain the datasets in the catalog that are linked to them.\footnote{\url{https://github.com/mclient-project/crawling/tree/main/completeness}}
Afterwards, the completeness score can be determined by calculating the ratio of datasets having all properties in the respective category among the total number of datasets. 
\begin{equation}
Completeness= \frac{\# datasets \; with \; mand.|rec. \; properties}{\# all \; datasets}
\end{equation}

\section{Use Case Analysis}
\label{sec:usecase}
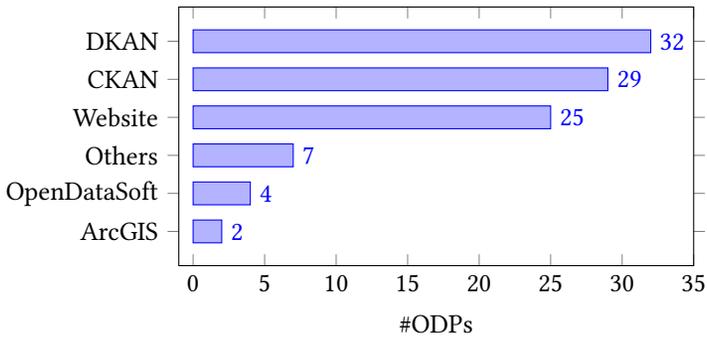
\begin{figure}
\begin{center}
\begin{tikzpicture}
\begin{axis}[
  xbar, 
  y=-0.5cm,
  bar width=0.3cm,
  enlarge y limits={abs=0.45cm},
  xlabel={\#ODPs},
  symbolic y coords={DKAN,CKAN,Website,Others,OpenDataSoft,ArcGIS},
  ytick=data,
  nodes near coords, nodes near coords align={horizontal},
  ]
\addplot table[col sep=comma,header=false] {
32,DKAN
29,CKAN
25,Website
7,Others
4,OpenDataSoft
2,ArcGIS
};
\end{axis}
\end{tikzpicture}
\end{center}
\caption{Open Data Portal Technologies }
\label{fig:odp}
\end{figure}

In order to validate our evaluation model, we applied it to the landscape of Open Data portals in Germany. The German ODL is comprised of portals of municipalities, counties, states and federal authorities that publish self-collected data. The legal foundation for these publishing activities is the German Open Data Act (EGovG) that was passed in 2017. The law requires public bodies in Germany to make their data sources publicly available if no other reasons speak against it. Data sources to be published are those that have been collected by the authorities in order to fulfill public tasks. The prerequisite is that the data is electronically stored by the authority and that it is available in processable formats, such as tables or lists.\footnote{\url{https://www.gesetze-im-internet.de/egovg/__12a.html}}\\
For the assessment of the German Open Data landscape, we utilized a curated list of all known German Open Data portals. This list can be obtained from a public CKAN instance of the city of Moers.\footnote{\url{https://www.offenesdatenportal.de/dataset/ubersicht-der-open-data-angebote-in-deutschland}}. The list comprises information of 178 Open Data portals in Germany. Only some ODPs of this list have a machine-readable interface, while many provide their data collections solely through a website (see Fig. \ref{fig:odp}). 

Only the ODPs with a public standardized API (e.g., CKAN or DKAN) were crawled for the use case analysis since a scraping of HTML sites can not be sustained for a large number of non-API websites. In total, we crawled 62 Open Data portals. Since almost half of these portals are part of a larger platform (e.g., at the state level), some of those identified were combined, resulting in a directory of 28 metadata catalogs as an outcome of the overall crawling process. The DCAT metadata descriptions serve as the foundation of our use case analysis and are publicly available on Github in \verb+turtle+ format and can thus be queried with a SPARQL-enabled database.\footnote{\url{https://github.com/mclient-project/crawling/tree/main/catalog/toLoad}} As the ODPs can be reached via different APIs and some even offer original DCAT descriptions, our approach to the creation of the DCAT registry was two-fold:
\begin{itemize}
\item \textit{ODPs with a DCAT API}: For those Open Data portals that already offer DCAT metadata descriptions, we followed the straightforward approach of subsequently downloading these files via sucessive \verb+CURL+ requests. We validated syntactic correctness with the command-line tool \verb+raptorutils+ and removed incorrect lines (e.g., such as faulty IRIs) until the validated output did not produce errors anymore. 
\item \textit{ODPs with a JSON API:} For ODPs only providing a standard JSON output, we utilized the \verb+dcat-suite+, a software application for easy publication, retrieval and conversion of dataset descriptions, to provide the metadata in DCAT format. In addition to the automatic format conversion, the \verb+dcat-suite+ is also able to correct potential errors in the data and to query interfaces provided by CKAN as well as DKAN instances.\footnote{\url{https://github.com/SmartDataAnalytics/dcat-suite/tree/develop}} 
\end{itemize}

The individual processing steps incl. post-processing operations (i.e., namespace replacements, adding to the overall meta-catalog registry) have been automated with the \verb+RDFProcessingToolkit+ and the \verb+Make+ tool. They can be executed as successive CLI commands. To update the DCAT metadata registry, the crawling pipeline can be repeated for the entire list of crawlable ODPs at any given time and takes currently up to 15 minutes to complete. The pipeline can also be easily adapted for the crawling of other Open Data portals and Open Data landscapes.\footnote{\url{https://github.com/mclient-project/crawling}}  

\subsection{Key Data}
Table \ref{tab:key} shows the number of DCAT datasets, distributions and access URLS in the German Open Data landscape. It also depicts the average number of entities in each of these categories for a single Open Data portal. The absolute number of datasets in the German ODL is quite high. However, compared to the number of datasets being published in other industrial nations, there is still room for improvement. In their worldwide benchmark of Open Data portals Neumaier et al. found out that there are individual Open Data portals in Canada and the USA that provide more than twice as many data collections as the entire Open Data landscape in Germany. They report that the central Canadian (\verb+www.data.gc.ca+) and US American (\verb+data.gov+) Open Data hubs facilitate access to more than 240,000 and more than 160,000 datasets respectively. Thus, it can be stated that Germany is still in an emerging stage in terms of Open Data adoption. An observation that is also confirmed by the Open Data barometer (ODP) \cite{odb}. However, looking at the average number of data collections per portal and taking into account comparative values from previous international benchmarks, it can be said that the individual data portals in Germany provide a comparatively large number of data collections \cite{braunschweig}.

\begin{table} [!htbp]
\centering
\label{tab:survey}
\begin{tabular}{cccc}
\toprule
 & \textbf{\#DCAT datasets} & \textbf{\#DCAT distributions} & \textbf{\#DCAT access URLs} \\
\midrule
All ODPs & 47,088 & 127,429 & 115,906 \\
Mean per ODP & 1624 & 4394 & 3997 \\
\bottomrule
\end{tabular}
\caption{Statistics on key data in the meta-catalog of German ODPs}
\label{tab:key}
\end{table}

\subsection{Topics}
\begin{figure}[h!]
    \centering
   \includegraphics[width=\textwidth]{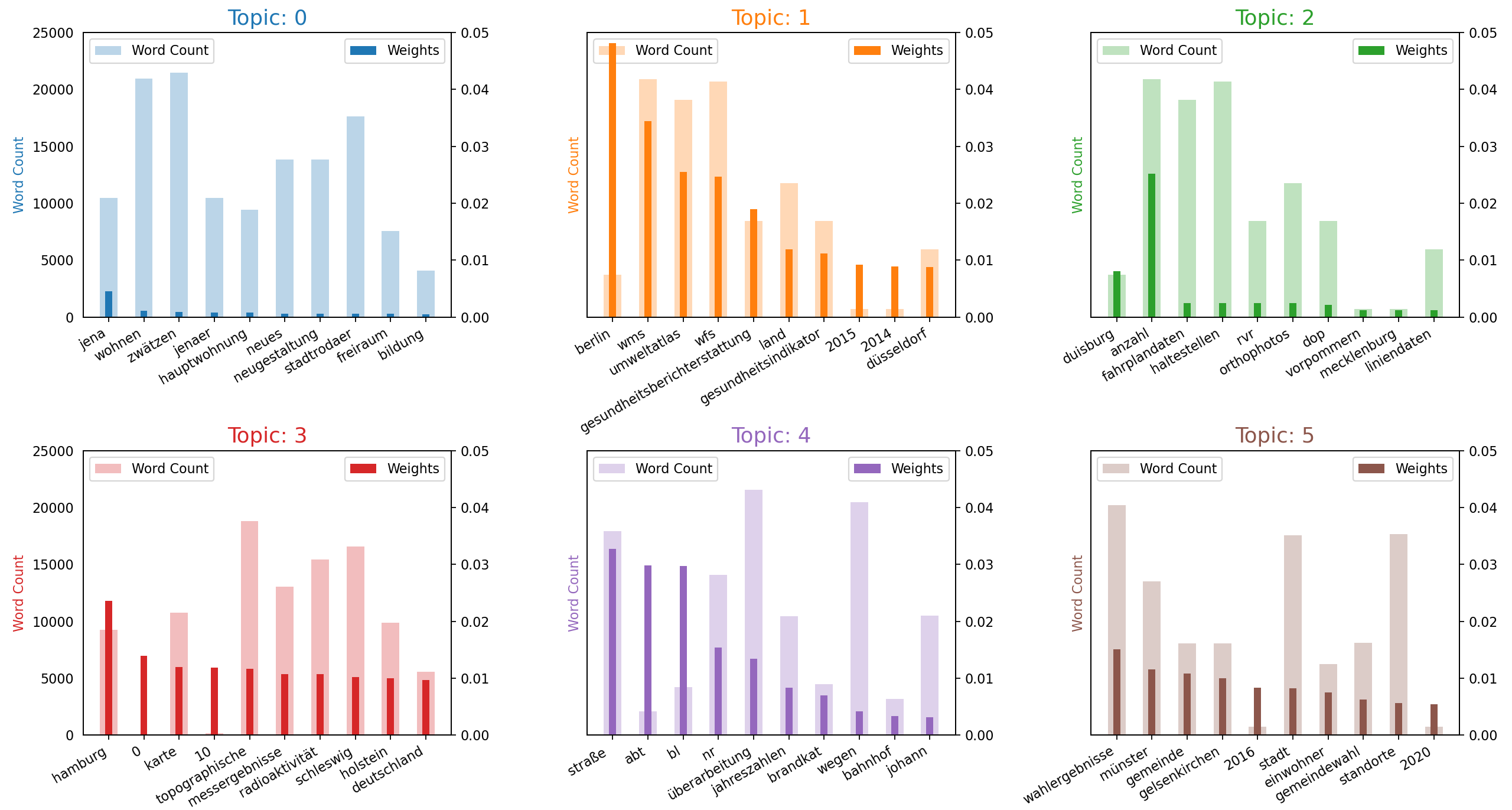}
    \caption{Identified topics and relative importance of related words}
    \label{fig:topic_hist}
\end{figure}

We conducted the topic analysis by utilizing the Python package \verb+sklearn+.\footnote{\url{https://scikit-learn.org/stable/modules/generated/sklearn.decomposition.LatentDirichletAllocation.html}} The package implements the \textit{Latent Dirichlet Allocation} (LDA) algorithm \cite{blei}. It can generate topic maps based on the text descriptions in a corpus. For each dataset, we extracted the corresponding titles (via the property \verb+dct:title+) from the generated catalog of DCAT metadata. These text descriptions were modified with a standard pipeline of NLP operations, such as tokenization and stopword removal. Afterwards, the LDA model was trained to fit the data. We experimented with the training configuration and reached a meaningful solution with 1000 training iterations and a final selection of six topic clusters.\\
Figure \ref{fig:topic_hist} depicts the results. The diagrams visualize the word counts (bold bars) for terms that are aligned on the x-axis and compares them to the relative importance of these words in the cluster as determined by their respective weights (faint bars). Table \ref{tab:key} lists a summarizing description for each topic cluster which was derived from the 10 most important terms in each group. The terms listed in the table are the English translation of the German words being shown in the diagrams. The translations were made to increase transparency and to make the manual assignment of topic descriptions comprehensible for the English reader. As can be seen from the table, there exists a thematic focus on geodata, public transport and local government issues (e.g., fire prevention and elections) in the German Open Data landscape. While this set of topics is already quite diverse, other topics, such as public health, social issues, education as well as business and economics may be of equal importance for stakeholders and should therefore be considered more by Open Data providers in the future. 

\begin{table} [!htbp]
\small
\centering
\label{tab:survey}
\begin{tabular}{p{0.05\textwidth}p{0.25\textwidth}p{0.55\textwidth}}
\toprule
\textbf{Topic} & \textbf{Description} & \textbf{Terms (English Translation)} \\
\midrule
0 & \textit{Housing} & \verb+jena+, \verb+housing+, \verb+zwätzen+, \verb+jena people+, \verb+main residence+, \verb+new+, \verb+redesign+, \verb+stadtroda people+, \verb+free space+, \verb+education+ \\
\hline
1 & \textit{Environment and Health} & \verb+berlin+, \verb+wms+, \verb+environmental atlas+, \verb+wfs+, \verb+environmental reporting+, \verb+state+, \verb+health indicator+, \verb+2015+, \verb+2016+, \verb+düsseldorf+\\
\hline
2 & \textit{Public Transport} & \verb+duisburg+, \verb+number+, \verb+timetable data+, \verb+stations+, \verb+rvr+, \verb+orthophotos+, \verb+dop+, \verb+pomerania+, \verb+mecklenburg+, \verb+schedule data+\\
\hline
3 & \textit{Map and Geodata} & \verb+hamburg+, \verb+0+, \verb+map+, \verb+10+, \verb+topographical+, \verb+measurement results+, \verb+radioactivity+, \verb+schleswig+, \verb+holstein+, \verb+germany+ \\
\hline
4 & \textit{Fire and Disaster Prevention} & \verb+street+, \verb+abt+, \verb+bl+, \verb+nr+, \verb+revision+, \verb+annals+, \verb+fire and disaster prevention+, \verb+due to+, \verb+station+, \verb+johann+ \\
\hline
5 & \textit{Elections} & \verb+election results+, \verb+münster+, \verb+municipality+, \verb+gelsenkirchen+, \verb+2016+, \verb+city+, \verb+inhabitants+, \verb+municipal election+, \verb+locations+, \verb+2020+\\
\bottomrule
\end{tabular}
\caption{Description of topics}
\label{tab:key}
\end{table}

\subsection{Freshness}
\begin{figure}
\begin{minipage}[t]{0.45\textwidth}
\includegraphics[width=\textwidth]{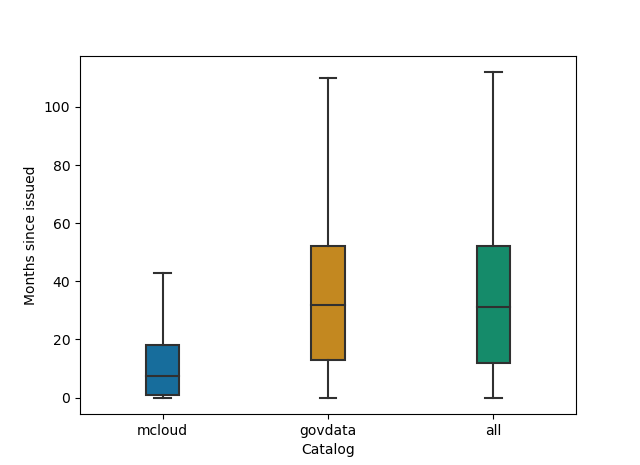}
\end{minipage}
\begin{minipage}[t]{0.45\textwidth}
\includegraphics[width=\textwidth]{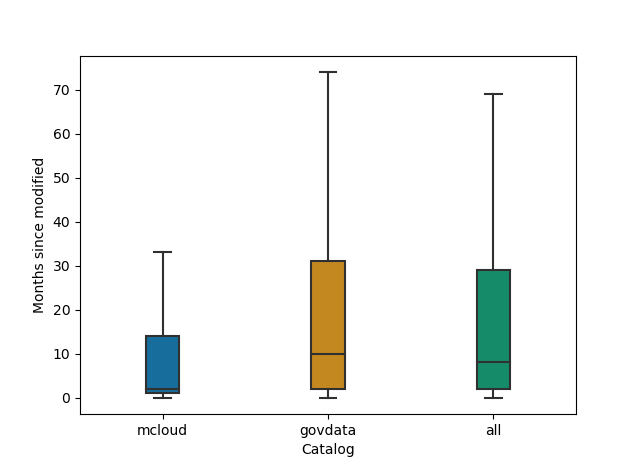}
\end{minipage}
\cprotect\caption{Months past since the publication (\verb+dct:issued+) and modification (\verb+dct:modified+) of a dataset}
\label{fig:timestamps}
\end{figure}

Freshness was assessed by analyzing the publication and modification timestamps attached to the metadata descriptions of the datasets. We extracted the values of the properties \verb+dct:issued+ and \verb+dct:modified+ and calculated the difference between the date of the latest crawl and the publication and modification timestamps respectively. Figure \ref{fig:timestamps} shows the distributions of freshness data (i.e., the number of months) for the two largest ODPs in Germany \verb+www.mcloud.de+ and \verb+www.govdata.de+ as well as for the entirety of all German Open Data portals (\verb+all+). The Federal Ministry of Transport and Digital Infrastructure (BMVI) operates the mCLOUD portal as a data hub for information on transportation, climate \& weather, aerospace, and general infrastructure. Govdata, on the other hand, is the central data hub for openly accessible datasets with no particular thematic focus. It can be seen that the mCLOUD portal - on average - provides access to datasets with a more recent publication date. The same holds true for the timestamps of the last modification date, since these are also more recent in the mCLOUD portal. This finding is not necessarily surprising as the mCLOUD only provides data collections on a clearly defined subject area and its catalog is therefore much smaller in size. Thus, the freshness scores for the govdata portal as well as for all portals show a larger range since they register considerably more datasets. As pointed out in Subsect. \ref{subsec:fresh} the timeliness dimension is closely linked to the actual needs of the end users. Hence, a pure analysis of the actuality of data collections does not allow for clear conclusions to be drawn about their quality. 

\subsection{Location Coverage}
We applied the NUTS hierarchy to capture the location coverage of the German ODL. The analysis was based on the list of German ODPs provided by the city of Moers, which stated the NUTS code for each Open Data portal. The NUTS hierarchy assigns government units to different administrative levels. In the German administrative system, NUTS level 1 is represented by the 16 federal states of the country. NUTS level 2 refers to the 29 current and former government districts and NUTS level 3 comprises the 401 counties/county districts and independent cities. Our analysis could match Open Data portals to NUTS levels 1 and 3, but not to NUTS level 2. This might be due to the fact that ODPs belonging to federal states also serve as data hubs for downstream administrative units. On the other hand, this does not mean that data collections from administrative units at level 3 cannot also be found in the central data hubs of the federal states (level 1) which might also explain the low coverage of level 3 as is shown in Table \ref{tab:nuts}. On level 1, half of the federal states in Germany already have an Open Data infrastructure in place.\\
Considering the fact that the German Open Data Act (EGovG) requires each administrative unit to make eligible data sources publicly available, the location coverage seems to be low. To mitigate this statement, however, it must be taken into account that besides data portals at the federal level, Germany's central data hub (\verb+govdata+) also offers the opportunity for public bodies to publish their data collections. For many administrative units not having their own platform this is the only option for making their data sources publicly available. Therefore, the coverage might be a bit higher than the numbers suggest. 
\begin{table*}[!htbp]
\centering
\label{tab:survey}
\begin{tabular}{ccc}
\toprule
 & \textbf{NUTS-Ratio$_1$} & \textbf{NUTS-Ratio$_3$} \\
\midrule
All ODPs & 0.5 (8/16) & 0.08 (32/401) \\
API-based ODPs & 0.25 (4/16) & 0.04 (15/401) \\
\bottomrule
\end{tabular}
\caption{Administrative units directly covered by Open Data Portals in Germany}
\label{tab:nuts}
\end{table*}

\subsection{Uniqueness}
We utilized the entire meta-catalog of the German ODL for uniqueness calculation. For each dataset and each one of the relevant properties (i.e., \verb+dct:identifier+, \verb+dct:title+, \verb+dct:description+), it was determined how often the corresponding values occurred in the catalog. Therefore, we made sure that dataset duplicates were left out of the analysis as they would have distorted uniqueness calculation. We only detected duplicate identifiers, titles or descriptions for otherwise \textit{different} datasets. The underlying assumption being that the more unique features are, the better the datasets can be found and distinguished from others.\\
Table \ref{tab:unique} shows the results of the calculation by listing the mean, standard deviation (\textit{Std}), maximum (\textit{Max}) and minimum scores (\textit{Min}) for each property and the aggregated compound score respectively. It becomes clear that item features are for the most part used quite unambiguously within the German ODL as is shown by the high numbers in the average uniqueness score categories for each property. Among them, dataset descriptions seem to be the most problematic property indicating that some datasets throughout the catalog are ambiguously described. This finding is in line with the assumption that there is still room for improvement when it comes to meaningful dataset descriptions. 
\begin{table*}[!htbp] 
\centering
\label{tab:survey}
\begin{tabular}{lcccc}
\toprule
& \verb+dct:identifier+ & \verb+dct:title+ & \verb+dct:description+ & \textbf{compound}\\
\midrule
Mean & 0.995 & 0.961 & 0.840 & 0.947 \\
Std & 0.155 & 0.110 & 0.268 & 0.071 \\
Max & 1.000 & 1.000 & 1.000 & 1.000 \\
Min & 0.917 & 0.484 & 0.252 & 0.660 \\
\bottomrule
\end{tabular}
\caption{Uniqueness scores for the different properties}
\label{tab:unique}
\end{table*}

\subsection{Interoperability}
\begin{figure}
\caption{Distribution of DCAT distribution among top file formats}
\label{fig:format}
\begin{tikzpicture}
\begin{axis}[
  xbar, 
  y=-0.5cm,
  bar width=0.1cm,
  enlarge y limits={abs=0.45cm},
  xmin=0,
  xmax=35000,
  xlabel={\#DCAT distributions},
  symbolic y coords={1.CSV,2.PDF,3.XML,4.JSON,5.HTML,6.XLSX,7.XLS,8.GML,9.GEOJSON,10.SHP,...,25.RDF},
  ytick=data,
  nodes near coords, nodes near coords align={horizontal},
  ]
\addplot table[col sep=comma,header=false] {
16745,1.CSV
4969,2.PDF
4117,3.XML
3181,4.JSON
3059,5.HTML
2806,6.XLSX
2133,7.XLS
1392,8.GML
843,9.GEOJSON
656,10.SHP
nan,...
45,25.RDF
};
\end{axis}
\end{tikzpicture}
\end{figure}
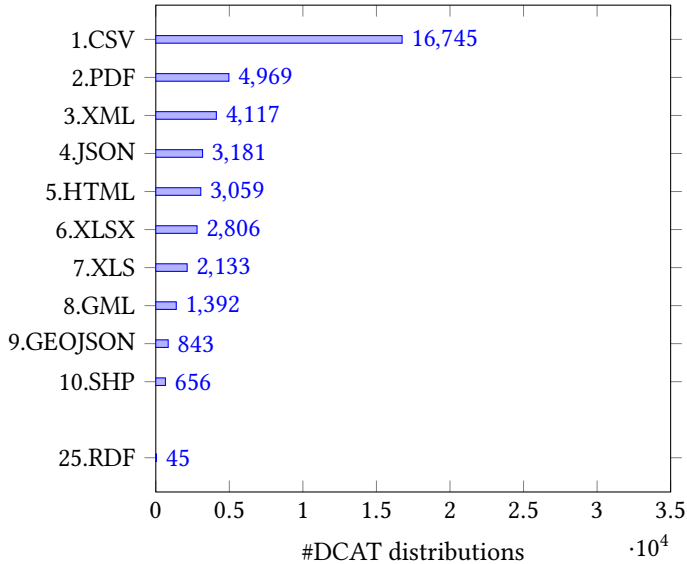

Table \ref{tab:inter} gives an overview of the interoperability scores. They were obtained by analyzing the technical features of German ODPs and by surveying the prevalence of types of file formats. In total, more than half of the Open Data portals in Germany provide an Open API (as measured by the \textit{Open Ratio}). Thus, software agents can access metadata descriptions as well as download datasets automatically from these portals. In the future, even more platforms should offer an open interface so that the processing of data catalogs can be carried out more efficiently. The same holds true for the \textit{DCAT Ratio} since only slightly more than 10\% of the surveyed portals provide DCAT information themselves. This means that the vast majority of metadata descriptions must first be converted from other formats, such as JSON, into DCAT in order to achieve a sufficient degree of interoperability. Since the transformation is error-prone, it would be better if portal owners provided DCAT metadata right from the start.\\
Things look better for the provision of data in open formats. File formats of DCAT distributions were determined from the crawled meta-catalog of German ODPs. There are 88\% of DCAT distributions available in open formats (\textit{Open-Format-Ratio}). This high percentage is a positive result. It shows that there is a strong awareness on the part of data owners and platform operators regarding the provision of open, non-proprietary data collections. In addition, Fig. \ref{fig:format} depicts the numbers of DCAT distributions among the top file formats which further illustrates the high prevalence of open file types, such as \verb+CSV+, \verb+XML+ or \verb+HTML+. It also shows that the vast majority of files are in semi-structured or unstructured text formats. \textit{Linked Open Formats}, such as RDF, only account for less than 0,1\% of the registered distributions in the catalog. 

\begin{table*}[!htbp] 
\centering
\label{tab:survey}
\begin{tabular}{ccc}
\toprule
\textbf{Open-Ratio} & \textbf{DCAT-Ratio} & \textbf{Open-Format-Ratio} \\
\midrule
0.61 & 0.13 & 0.88 \\
\bottomrule
\end{tabular}
\caption{Overview of Interoperability scores}
\label{tab:inter}
\end{table*}

This finding is in line with previous research on file type prevalence in dataset registries. For instance, an analysis of the Google dataset search catalog revealed that Linked Data files make up less than 1\% of the worldwide dataset landscape. Against the background of data profiling analyses, the authors assume that in all likelihood there exist a larger number of Linked Data collections, but that these are also insufficiently described by metadata, so that they are often not found \cite{numbers}. In the case of ODPs, another complicating factor is that many public institutions do not have the appropriate technical and personnel capacities to convert data into interoperable formats. Even if not all data is suitable for these kinds of transformations, reusability could be alleviated in many cases.\\
The degree of interoperability within the metadata descriptions (\textit{Linked Metadata}) was also assessed. We obtained the results listed in Tab. \ref{tab:namespace} by issuing a corresponding SPARQL query (see Subsect. \ref{subsec:inter}). The table shows that - apart from the vocabularies for general entity description (e.g., \verb+rdf+, \verb+rdfs+, \verb+owl+, \verb+dct+) as well as for characterizing data and software artifacts (\verb+dcat+, \verb+spdx+, \verb+adms+) - hardly any external vocabularies are embedded in the catalog. Rare exceptions are the \verb+foaf+ vocabulary as well as the \verb+mdr+ and \verb+opd+ schemas which are used for assigning topic themes from a controlled vocabulary. Taking into account the fact that dataset descriptions should adhere to the DCAT standard, a more widespread use of Linked Data vocabularies would still be desirable to better integrate datasets into the Semantic Web infrastructure. 
\begin{table*}[!htbp] 
\centering
\small
\label{tab:survey}
\begin{tabular}{lll}
\toprule
\textbf{Abbreviation} & \textbf{Namespace} & \textbf{Occurrences} \\
\midrule
\verb+dcat+ & \verb+http://www.w3.org/ns/dcat#+ & 1,212,451 \\
\verb+dct+ & \verb+http://purl.org/dc/terms/+ & 1,006,223 \\
\verb+rdf+ & \verb+http://www.w3.org/1999/02/22-rdf-syntax-ns#+ & 236,330 \\
\verb+dcatde+ & \verb+http://dcat-ap.de/def/dcatde/+ & 149,971 \\
\verb+foaf+ & \verb+http://xmlns.com/foaf/0.1/+ & 105438 \\
\verb+mdr+ & \verb+http://publications.europa.eu/resource/authority/data-theme/+ & 85362 \\
\verb+spdx+ & \verb+http://spdx.org/rdf/terms#+ & 22,762 \\
\verb+adms+ & \verb+http://www.w3.org/ns/adms#+ & 14,654 \\
\verb+owl+ & \verb+http://www.w3.org/2002/07/owl#+ & 9697 \\
\verb+dcatde101+ & \verb+http://dcat-ap.de/def/dcatde/1.0.1/+ & 5300 \\
\verb+dcatde11+ & \verb+https://www.dcat-ap.de/def/dcatde/1.1/+ & 1281\\
\verb+dcatde10+ & \verb+http://dcat-ap.de/def/dcatde/1_0/+ & 397\\
\verb+opd+ & \verb+https://opendata.potsdam.de/id/theme/+ & 64\\
\verb+rdfs+ & \verb+http://www.w3.org/2000/01/rdf-schema#+ & 1\\
\bottomrule
\end{tabular}
\caption{Distribution of namespaces in the metadata catalog}
\label{tab:namespace}
\end{table*}

\subsection{Legal Security \& Openness}
In order to assess the legal security of the German ODL, we counted the number of licenses that are linked to DCAT distributions in the catalog. By that, we obtained the \textit{License Ratio} and found out that more than 90\% of distributions have a license statement in their metadata description (see \ref{tab:license}). Among them, the URL-based license statements, such as the Data license Germany\footnote{\url{https://www.govdata.de/dl-de/by-2-0}} or the Creative Commons (\verb+CC BY 4.0+) license\footnote{\url{https://creativecommons.org/licenses/by/4.0/}} were each referencing an Open license. We verified the openness of a license through cross-checking with the list of open licenses provided by the \verb+DCAT-AP.de+ platform.\footnote{\url{https://www.dcat-ap.de/def/licenses/}} In total, the \textit{Open License Ratio} amounted to 88\% of DCAT distributions. On the other hand, the text-based license statements were not particularly meaningful. Expressions, such as \textit{None} or \textit{Not specified} or cryptic hash values do provide zero information on a dataset's terms of use.

\begin{table*}[!htbp] 
\centering
\label{tab:survey}
\begin{tabular}{cc}
\toprule
\textbf{License Ratio} & \textbf{Open License Ratio} \\
\midrule
0.902 & 0.888 \\
\bottomrule
\end{tabular}
\caption{Prevalence of license statements among DCAT distributions}
\label{tab:license}
\end{table*} 

\subsection{Findability \& Accessibility}
The \textit{Replica Ratio} was determined by issuing the SPARQL query of Subsect. \ref{subsec:find} against the DCAT catalog of German ODPs. Table \ref{tab:find} shows that 2,8 \% of datasets and their corresponding \verb+dcat:accessURLs+ are listed in more than one Open Data portal. The analysis also revealed that the majority of replicas are copies of a dataset from a municipal ODP that have also been registered in Germany's central data hubs \verb+mCLOUD+ and \verb+govdata+. This replica rate is reasonable. It can be assumed that the registration of data collections in several catalogs increases their discoverability. in this respect, it could also be a bit higher. On the other hand, a data collection should always fit the thematic focus of the ODP it is listed in and should therefore not be registered on arbitrary platforms. 

\begin{table*}[!htbp] 
\label{tab:survey}
\begin{tabular}{ccc}
\toprule
\textbf{Replica Ratio} & \textbf{Mean Keyword IC} &  \textbf{Accessibility}\\
\midrule
0.028 & 0.410 & 0.369\\
\bottomrule
\end{tabular}
\caption{Findability and accessibility scores of the German ODL}
\label{tab:find}
\end{table*} 

\begin{figure}
    \centering
   \includegraphics[width=0.5\textwidth]{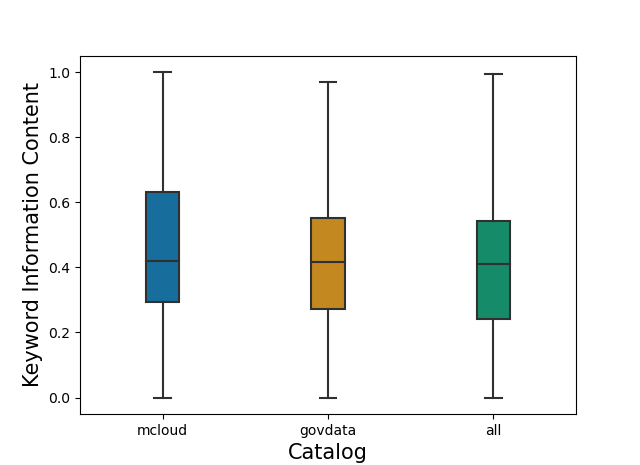}
    \caption{Distribution of \textit{Keyword Information Content} scores in different catalogs}
    \label{fig:ic}
\end{figure}

\begin{figure}
\begin{tikzpicture}
\begin{axis}[%
width=2in,
height=3in,
axis x line=center,
axis y line=left,
symbolic x coords={200,4xx,5xx},
enlargelimits=true,
enlarge x limits={abs=0.65cm},
ymin=0,
nodes near coords,
ylabel style={align=center},
ylabel={HTTP Status Code Counts},
xtick=data,
x tick label style={font=\small,align=center},
ybar]
\addplot coordinates {(5xx,1074) (4xx,2235) (200,47039)  };
\end{axis}
\end{tikzpicture}
\caption{Overview of HTTP status codes}
\label{fig:status}
\end{figure}
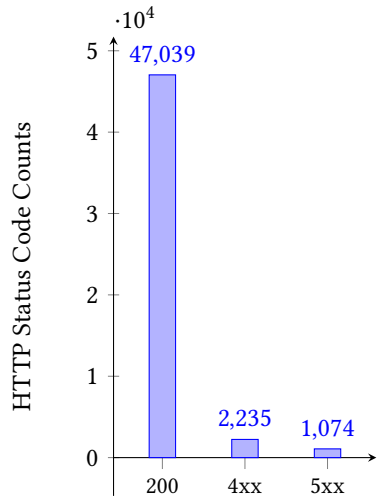

Another factor that increases the discoverability of data collections are thematic keywords. The degree to which datasets are described with meaningful topic words was measured with the \textit{Keyword Information Content (KIC)} metric. It determines the average information content of all keywords that are linked to a dataset description. Figure \ref{fig:ic} shows the distribution of \textit{KIC} scores throughout the German ODL (\verb+all+) and among the two largest German ODPs \verb+mcloud+ and \verb+govdata+. From the depicted boxplots, it can be concluded that, in most cases, topic descriptions have a low to medium information content score. It is also noteworthy that the distribution of the score values does not differ significantly between the individual data portals or portal landscapes. From these results, it can be argued that there is a considerable need for improvement with regard to a more expressive description of data collections.\\ 
In their analysis of the usage statistics of open research data, Quarati and Raffaghelli have shown that the use of datasets follows a steep power law distribution, with very high usage quantities for a small percentage of collections on the one hand and little to no downloads for the vast majority of data collections on the other \cite{researchers}. It seems reasonable to assume that an improved data description with meaningful thematic keywords could counterbalance these effects and would therefore also help interested citizens to find the datasets that meet their information needs.\\
The third category examined in this quality dimension was the aspect of \textit{accessibility}. To assess how easily users or software agents can actually access the data collections listed in Open Data portals, the \verb+dcat:accessURLs+ attached to the distributions were examined. The access URLs amounted to more than 100,000 links that were registered in the meta-catalog of German ODPs. For each one of these URLs, we issued a HTTP request and analyzed the status code of the corresponding HTTP response. Due to the high number of requests we had to make, we automated this step using python threading so that multiple web servers could be analyzed concurrently.\footnote{\url{https://github.com/mclient-project/crawling/tree/main/accessibility}}. In this way, data collection for the accessibility study was completed within a few minutes. The study revealed that only 36.9\% of DCAT distributions are linked to a working access URL (i.e., HTTP status code 200). For the remaining URLs, a corresponding HTTP request either resulted in an error code (see. Fig. \ref{fig:status}) or did not produce a response at all. Among those responses with an error code, client side errors (i.e., HTTP status codes 4xx) were the majority. They were most often caused by malformed URLs and missing access permissions. In the former case, errors might have often been caused by a webservice API (e.g., WMS or WFS) requiring additional parameters in order to successfully process the request. Hence, some of the services with error code responses will function properly provided that a user or software agent applies the appropriate query parameters. In addition, some server-side errors occurred (i.e., HTTP status codes 5xx), which in most cases can probably be explained by the fact that the corresponding service was down.\\
In summary, the accessibility of data sources leaves much to be desired. In the future, ODP operators should ensure that more data collections can actually be downloaded or they should add appropriate notes in the metadata if this is not possible. 

\subsection{DCAT Completeness}
For the assessment of the quality dimension \textit{DCAT completeness}, corresponding SPARQL queries were sent to the metadata catalog. In the category of mandatory properties, it was found that 97.5\% and thus almost all data collections have the mandatory property values of the DCAT-AP specification. This is a positive result. It shows that the majority of Open Data portals at least provide essential dataset properties such as the title and description of a dataset. In the category of recommended DCAT properties, the situation is somewhat different. Here, only about 25.6\% of the data collections available in the entire German Open Data landscape are described with all recommended properties (such as \verb+dcat:keyword+, \verb+dcat:contactPoint+ or \verb+dct:publisher+). Thus, it can be stated that ODP providers still need to catch up in terms of comprehensively describing datasets as well as in terms of the implementation of the DCAT standard. 

\section{Discussion} 
In this work, we presented a fine-grained framework for the evaluation of data and metadata quality in Open Data landscapes. Despite the growing adoption and implementation of Open Data strategies in countries throughout the world, there are currently still major gaps in our knowledge about whether the data collections listed in ODPs are actually described comprehensively and can thus be easily found. Such information is important in order to give appropriate feedback to Open Data operators regarding the quality of the metadata they provide. On top of that, data search engines, such as \textit{Google Dataset Search} or \textit{Figshare} also rely on good quality metadata in order to properly index resources and serve as additional access gateways to these data sources. However, a comprehensive literature survey showed that the vast majority of existing Open Data benchmarks currently either focuses on Open Data policies or only compares rudimentary indicators regarding DCAT completeness and conformance. Existing benchmarks thus lack a holistic approach towards measuring (meta)data quality. Therefore, building on results from related research areas such as Semantic Web, NLP and information retrieval, we have developed a comprehensive evaluation framework that captures all relevant aspects of Open Data quality, such as coverage, interoperability and findability. This evaluation framework is applicable to both individual Open Data portals and entire Open Data landscapes. It is thus also suitable for comparing multiple countries in terms of Open Data quality.\\
We have conducted an extensive analysis of the German Open Data landscape and also identified comparative quality scores for the largest Open Data portals in Germany. Our analysis has shown that the German Open Data landscape is still emerging. Numerous data collections are already available, which are also largely described conforming to the international DCAT standard. However, there are still deficits in many areas of metadata quality. For example, it would be desirable for existing descriptions to be more connected with the Linked Open Data cloud, e.g. by using internationally known keyword descriptors or by applying \verb+owl:sameAs+ links for the labeling and networking of data descriptions (quality dimension \textit{Interoperability}). Another area for improvement is the discoverability of data collections. Here, our analyses have shown that dataset descriptions are in some cases not yet unambiguous enough (\textit{Uniqueness}) and, in particular, that the expressive power of thematic keywords is rather limited. Hence, data collections might not be found by interested users (\textit{Findability}) and even if they are found, our research shows that the majority of data collections can not be accessed by the officially provided \verb+dcat:accessURL+. In the future, these effects should be mitigated by appropriate software tools for (semi-)automatic metadata enrichment as well as quality assurance for DCAT descriptions.\\
In summary, our fine-grained evaluation framework is able to provide an extensive picture of Open Data platforms through its scope as well as dedicated quantitative metrics. This has been validated by the analysis of the German Open Data landscape. The approach can be transferred to other platforms and landscapes for a comprehensive benchmarking of Open Data quality.

\begin{acks}
This work has been supported by the Federal Ministry of Transport and Digital Infrastructure (BMVI) under the grant number 19F2152A.
\end{acks}

\bibliographystyle{ACM-Reference-Format}
\bibliography{bibliography}

\end{document}